\documentclass[authoryear]{elsarticle}
\usepackage{mathrsfs}
\usepackage{pdflscape}
\usepackage{graphicx}
\usepackage{algorithmic}
\usepackage{algorithm}
\usepackage{listings}
\usepackage{amssymb}
\usepackage{amsmath}
\usepackage{amsfonts}
\usepackage{natbib}
\usepackage{lineno}
\usepackage{bm}
\usepackage{xr}

\newcommand{\be}{\begin{equation}} \newcommand{\ee}{\end{equation}}
\newcommand{\bd}{\begin{displaymath}} \newcommand{\ed}{\end{displaymath}}
\newcommand{\ba}{\begin{align}} \newcommand{\ea}{\end{align}}
\newcommand{\baa}{\begin{align*}} \newcommand{\eaa}{\end{align*}}
\newcommand{\ben}{\begin{enumerate}} \newcommand{\een}{\end{enumerate}}
\newcommand{\bi}{\begin{itemize}} \newcommand{\ei}{\end{itemize}}

\newcommand{\ud}{\mathrm{d}}
\newcommand{\E}[1]{\operatorname{E}\left[ #1 \right]}
\newcommand{\Var}[1]{\operatorname{Var}\left[ #1 \right]}
\newcommand{\var}[1]{\operatorname{Var}\left[ #1 \right]}

\newcommand{\Cov}[1]{\operatorname{Cov}\left[ #1 \right]}

\journal{Journal of Theoretical Biology}

\begin{document}

\begin{frontmatter}

\title{Interspecies correlation for neutrally evolving traits}

\author[CTHGU]{Serik Sagitov}
\ead{serik@chalmers.se}

\author[CTHGU]{Krzysztof Bartoszek\corref{cor1}}
\ead{krzbar@chalmers.se}

\cortext[cor1]{Corresponding author}
\address[CTHGU]{Mathematical Sciences, Chalmers University of Technology and the University of Gothenburg, Gothenburg, Sweden, Tel: +46 (0)31 772 10 00; fax: +46 (0)31-16 19 73}

\begin{abstract}
A simple way to model phenotypic evolution is to assume that after splitting, the trait values of the sister species diverge as 
independent Brownian motions. Relying only on a prior distribution for the underlying species tree (conditioned on the number, $n$, of extant species) we study the random vector $(X_1,\ldots,X_n)$ of the observed trait values. In this paper we derive compact formulae for the variance of the sample mean and the mean of the sample variance for the vector $(X_1,\ldots,X_n)$.
 
The key ingredient of these formulae is the correlation coefficient between two trait values randomly chosen from $(X_1,\ldots,X_n)$. 
This interspecies correlation coefficient takes into account not only variation due to the random sampling of two species out of $n$ and the 
stochastic nature of Brownian motion but also the uncertainty in the phylogenetic tree.  
The latter is modeled by a (supercritical or critical) conditioned branching process. In the critical case we modify the 
Aldous--Popovic model by assuming a proper prior for the time of origin.
\end{abstract}


\end{frontmatter}

\section{Introduction}
A simple way to model phenotypic evolution for $n$ related species is to assume that after splitting, the trait values  
(e.g. the logarithms of body sizes) of the sister species diverge as independent Brownian motions \citep[see][]{JFel85}. 
The resulting collection $(X_1,\ldots,X_n)$ of the tip species' trait values has a dependence structure 
caused by shared phylogeny. In this paper we derive compact formulae for the variance of the sample mean 
$\bar X=n^{-1}(X_1+\ldots+X_n)$ and the mean of the sample variance 
$S^2=(n-1)^{-1}\sum_{i=1}^n(X_i-\bar X)^2$. 
These formulae take into account not only the stochastic nature of 
Brownian motion but also uncertainty in the phylogenetic tree.

Based on observed tip species data one would like to make statements about the stochastic process
of evolution like  the ancestral state $X_0$ at the time of origin $T$ and infinitesimal variance 
$\sigma^2$ of the Brownian motion. These are
important questions addressed by phylogenetic comparative methods. Usually this sort of inference attempts to incorporate the knowledge
of the phylogenetic tree estimated from independent data
\citep{MButAKinOUCH,THanJPieSOrzSLOUCH,KBaretal}. There is however
uncertainty attached to the estimated tree which should be somehow reflected in
any subsequent analysis. 

All currently available methods addressing such statistical issues rely on simulations.
\citet{MPagFLut} and \citet{JHueBRan} propose to use an MCMC approach to generate a sample
of plausible phylogenetic trees each one with its posterior probability attached
as a weight. 
\citet{MButAKinOUCH} do not include phylogeny uncertainty
in their OUCH R \citep{R} package but say that in can be incorporated in their framework,
if one can compute likelihood values (e.g. posterior probabilities from a Bayesian estimation
procedure) for candidate trees. Then the complete likelihood function is a product of the 
tree's likelihood and the likelihood conditional on the tree and comparative data. A Bayesian estimation procedure implemented by
\citet{PLemARamJWelMSuc} uses a tree rescaling step,
with each branch of the phylogeny being independently rescaled by an appropriately (e.g. gamma or log--normal)
distributed random variable. 

These methods face a number of common challenges. The first one is computational, as estimating a
phylogeny can be computationally extremely demanding. The second is interpretational,
whilst the weighing of results is fully justified statistically one could
raise biological objections whether the result is actually biologically
meaningful for all parameters of the assumed model of trait evolution.
An extreme hypothetical example is if we would have two competing phylogenies
each with equal likelihood. The first results in a regression slope
of $1$, the second $-1$. The average of them is $0$. A regression slope 
of $0$ means that there is no relationship between the two variables
while both phylogenies indicate that there is a relationship
except that we don't have strong enough evolutionary data to decide 
about the direction of this relationship. The third problem
is that since we are merely ``trying out'' different possible
phylogenies we always run the risk of not considering the ones
close to the true one. 

Here we propose a different approach making use of explicit analytical calculations. We model the unknown phylogenetic tree for $n$ extant species using a conditioned birth--death process with 
speciation rate $\lambda$ and extinction rate $\mu$ as described by \citet{TGer2008}. 
The corresponding distribution of random trees with $n$ tips is a posterior distribution resulting 
from the improper uniform prior on the time of origin $T$. 
The appropriate range of the rates  $0\le\mu\le\lambda$ has an important region $\mu=\lambda$ 
representing the critical case \citep{DAldLPop} with the speciation and extinction events being equally likely. 
In the supercritical case $\mu<\lambda$ the height of the tree is expected to be lower due to the expansive speciation regime. 
A key test example of the supercritical birth--death model is the classical Yule model \citep{GYul1924} of pure birth process when $\mu=0$. 

\begin{figure}
\centering
\includegraphics[width=13cm]{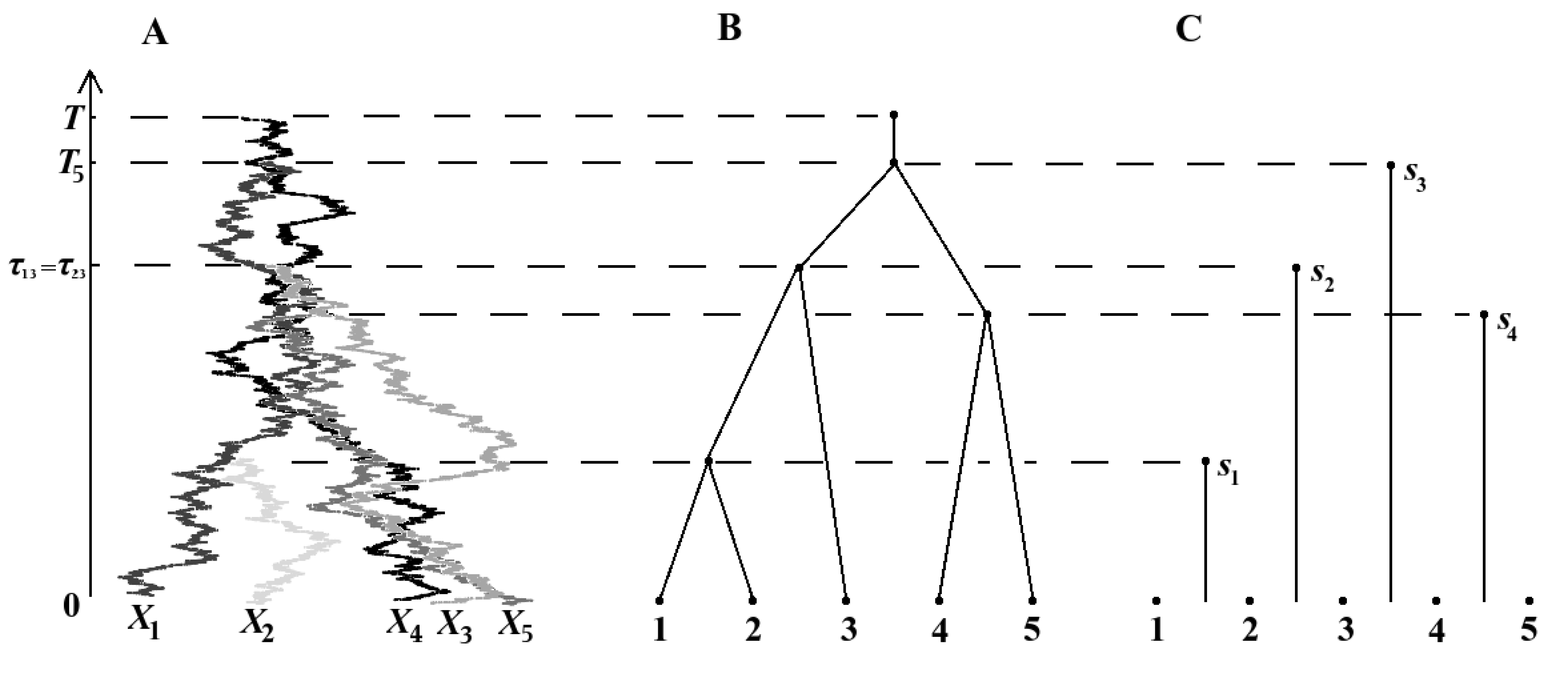}
\caption{A branching Brownian motion simulated on a random tree with $n=5$ tips using the TreeSim and mvSLOUCH software. 
Panel A:  the trait evolution for five species is modeled by a Brownian motion with $\sigma=1$. 
Panel B: the species tree disregarding the trait values. Panel C: a convenient presentation of speciation times.}
\label{tr}
\end{figure}

\section{Summary of main results}
In our setting both the variance of the sample mean
\be\label{eqVarXnBM}
\var{\bar{X}_{n}} = \sigma^{2}n^{-1}(1+(n-1)\rho_{n})\E{T}
\ee
and the mean of the sample variance
\be \label{eqES2BM}
\E{S^{2}_{n}}= \sigma^{2}(1-\rho_{n})\E{T},
\ee
are compactly expressed (see \ref{appA}) in terms of the correlation coefficient
\be\label{defrho}
\rho_{n} = \frac{1}{\binom{n}{2}\var{X} } \sum\limits_{1\le i< j\le n}\Cov{X_{i},X_{j}}
\ee
and the mean time to the origin $\E{T}$.

Sections \ref{sy},  \ref{ss},  \ref{sc} present analytical formulae for $\rho_n$ and $\E{T}$ 
in the Yule, supercritical and critical cases. These formulae are summarized in Tab. \ref{tbMainRes}
\begin{table}
\begin{center}
\begin{tabular}{|l|l|l|l|l|l|}\hline
Species tree       &Extinc-&Speciation&Exact &Approxi-&$\E{T}$\\
model &tion rate &rate &$\rho_n$&mate $\rho_n$&\\ \hline
Yule&$\mu=0$ & $\lambda=1$&\eqref{yf} & \eqref{ln}& \eqref{yt}\\
supercritical&$\mu=1$ & $\lambda>1$&\eqref{sf} & \eqref{ln1}& \eqref{st}\\
near-critical&$\mu=1$ & $\lambda>1$, $\lambda\approx1$&\eqref{sf} & \eqref{cf}, \eqref{cfa}& \eqref{st}\\
PP-critical&$\mu=1$ & $\lambda=1$&\eqref{cf1} & \eqref{cf2}, \eqref{cfa1}& \eqref{ct}\\\hline
 \end{tabular}
\end{center}
\caption{Summary of formulae for $\rho_n$ and 
$\E{T}$.}\label{tbMainRes}
\end{table}
in terms of three principal cases for the species tree model.
Observe that we incur no loss of generality by specifying one of the two parameters $(\mu,\lambda)$. For example, 
in a seemingly more general case with $0<\mu<\lambda$ the same formula, Eq. \eqref{sf} holds with $\lambda$ replaced by the ratio $\lambda/\mu$. 

\begin{figure}
\begin{center}
\includegraphics[width=0.45\textwidth]{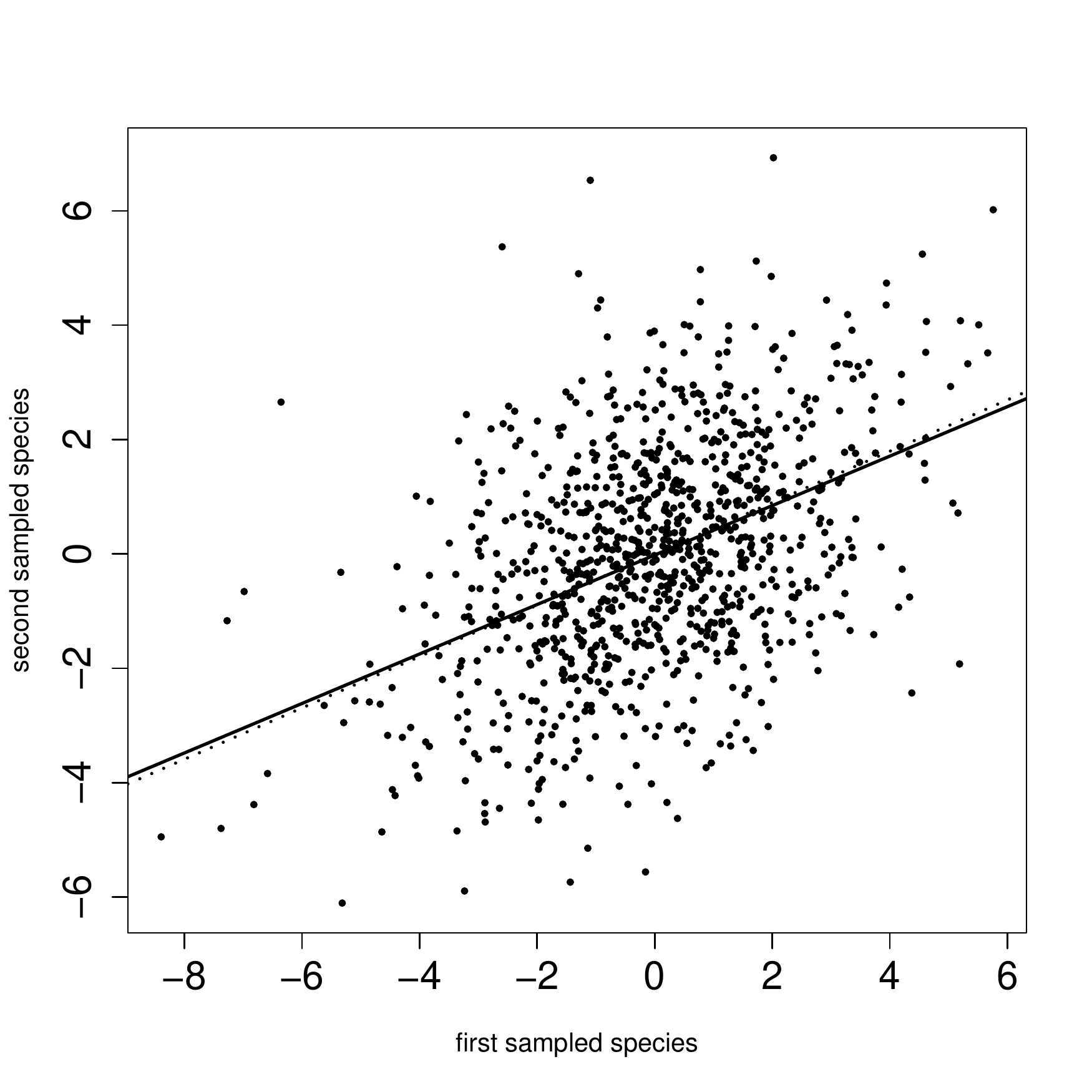}
\includegraphics[width=0.45\textwidth]{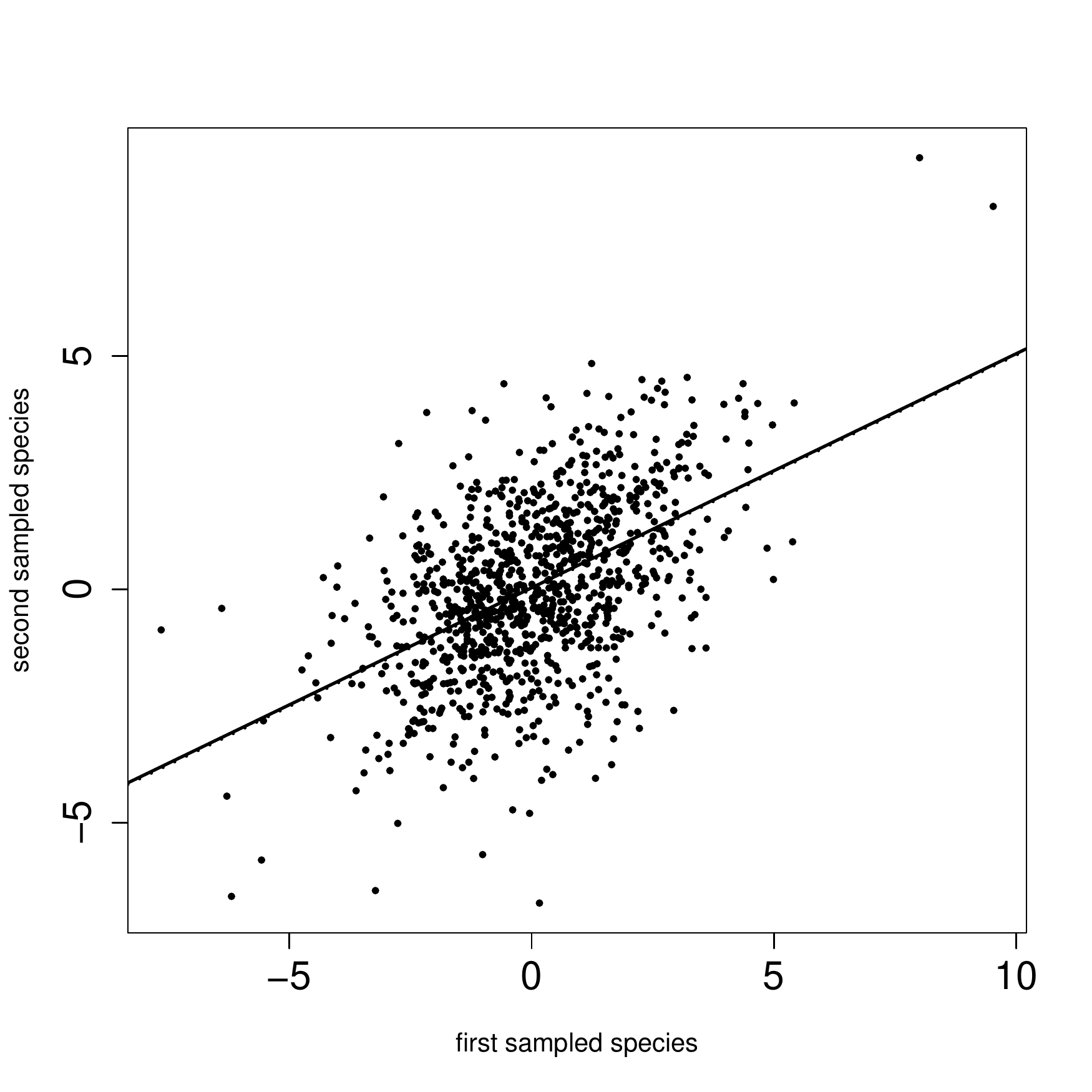} 
\includegraphics[width=0.45\textwidth]{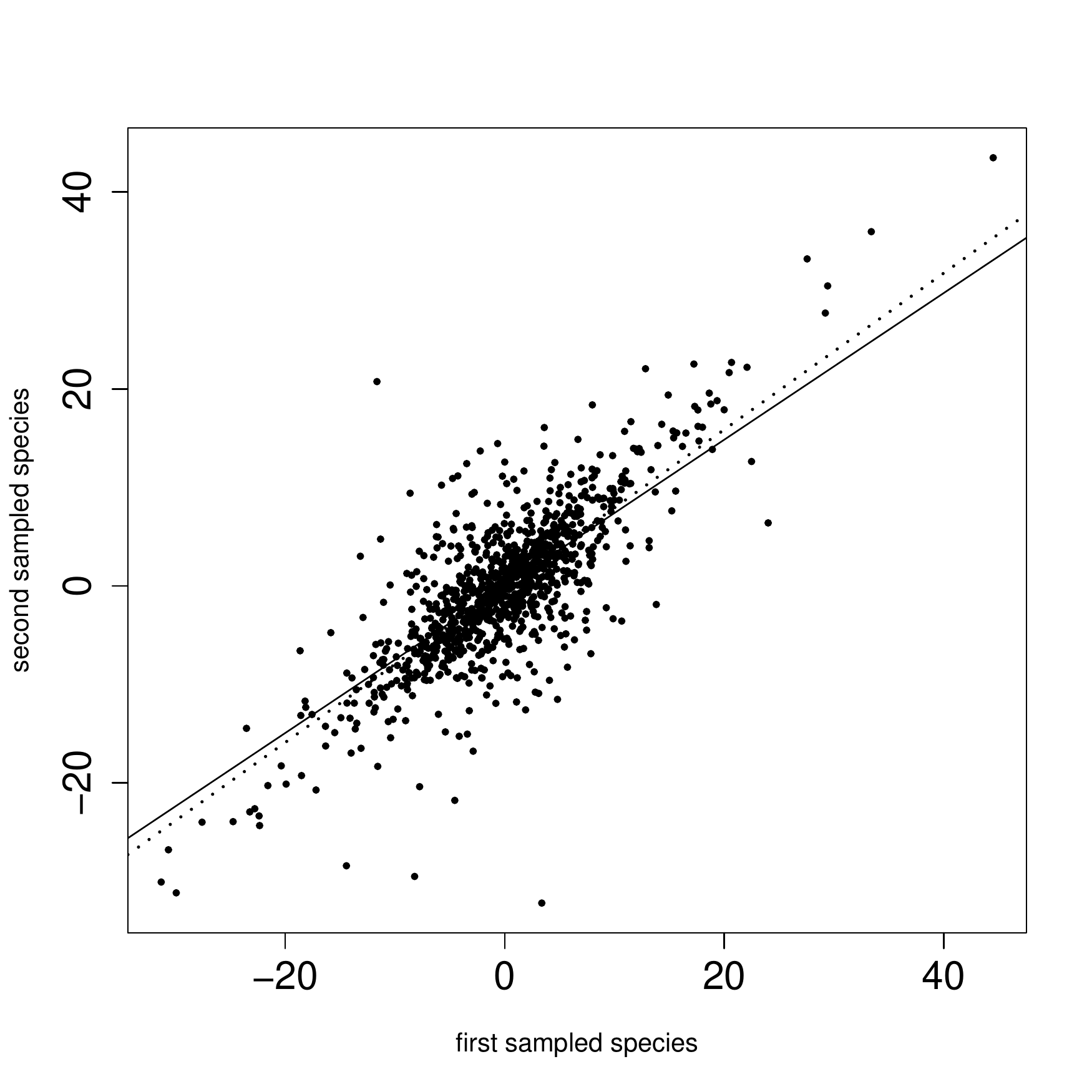}
\includegraphics[width=0.45\textwidth]{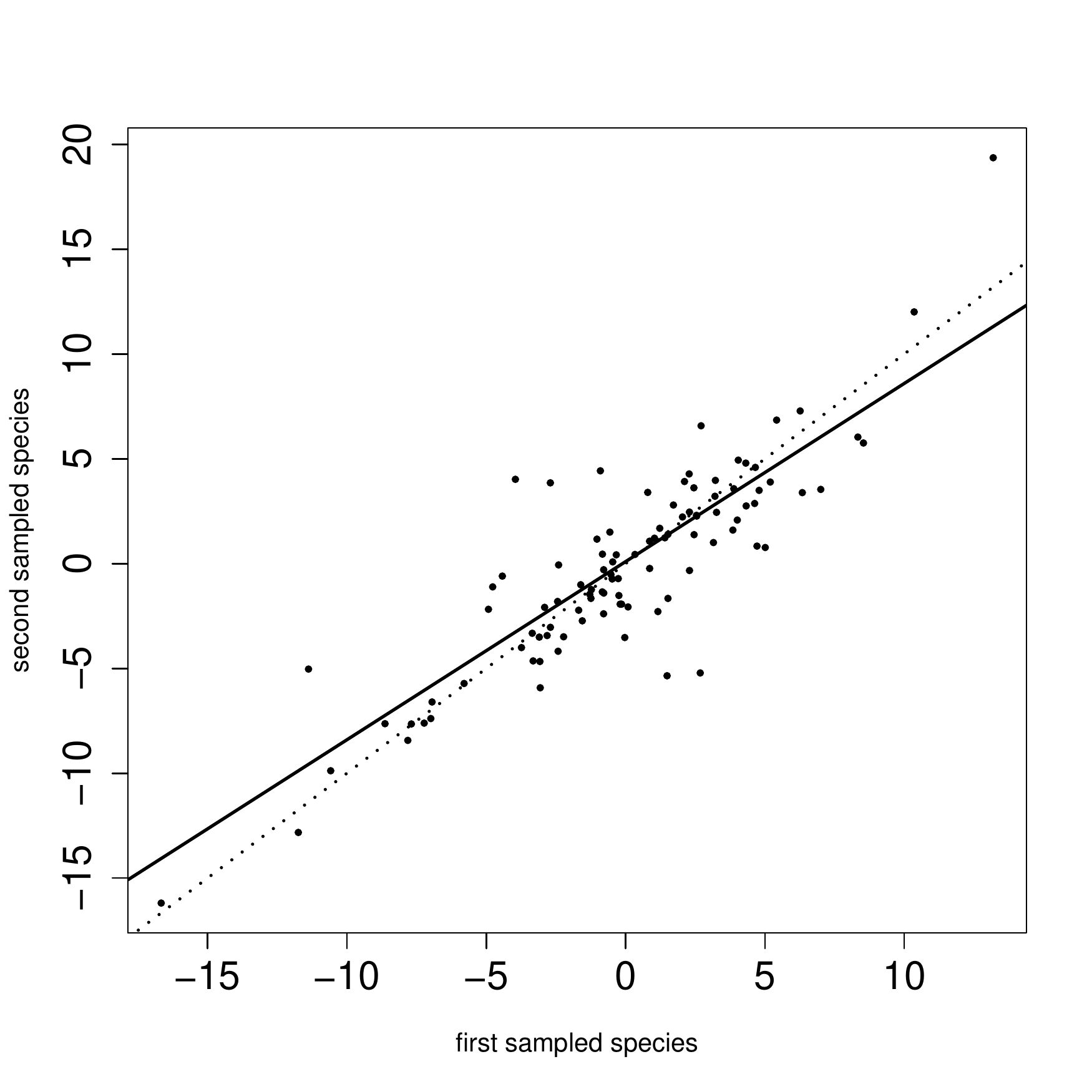}
\end{center}
\caption{Regression line fitted to the simulated data (thick line) 
compared to the line $y=\rho_n x$ (dotted line) with $\rho_n$ given by the exact formula. 
Upper--left panel: the Yule case. Upper--right: the supercritical 
case with $\lambda=2$. Lower--left: the near--critical case with $\lambda=1.01$. 
Lower--right: the critical case with improper prior (here the dotted line is $y=x$).}
\label{f4}
\end{figure}

What we call the interspecies correlation coefficient $\rho_n$ is the correlation between two trait 
values randomly chosen among  $n$ observed. 
Next, to clarify the exact meaning of  $\rho_n$ we describe an algorithm producing a pair of random variables having $\rho_n$ as the correlation 
coefficient for a given set of parameters $(n,\lambda)$.
\begin{algorithm}[!h]
\caption{Generate 
two random variables with a correlation of $\rho_{n}$ }\label{algX1X2}
\begin{algorithmic}[1]
\STATE generate a species tree with $n$ tips using TreeSim \citep{TreeSim1,TreeSim2},
\STATE generate $n$ trait values by running a branching Brownian motion over the species tree simulated in step 1 using mvSLOUCH \citep{KBaretal},
\STATE choose at random two out of $n$ trait values generated in step 2.
\end{algorithmic}
\end{algorithm}
The steps 1--3 of Algorithm \ref{algX1X2} (implemented by us in R) were repeated many times to collect enough data for estimating the 
correlation coefficient between the underlying pair of random variables, see Fig. \ref{f4}. 
The simulation results presented in Tab. \ref{tbSimRes}
compare the correlation coefficient estimated from the simulated trees $\hat\rho_n$ to the true value of $\rho_n$ 
and the value given by an appropriate approximate formula. 
Notice that we did not simulate the critical case with a proper prior as suitable software is currently lacking. Simulations of the critical case with improper prior are time consuming. Therefore the critical case is represented with a smaller number of dots on the graph.

\begin{table}
\begin{center}
\begin{tabular}{|l|ccccc|}\hline
Model & $n$ & Trees  & $\hat\rho_n$ & $\rho_n$& Approximation \\
\hline
$\mu=0$,  $\lambda=1$ & $30$ & $1000$  & $0.430$& $0.449$ & $0.503$ using \eqref{ln} \\
$\mu=1$,  $\lambda=2$ & $30$ & $1000$  & $0.506$& $0.502$ & $0.609$ using \eqref{ln1} \\
$\mu=1$,  $\lambda=1.01$ & $30$ & $1000$ & $0.784$ & $0.794$ & $0.689$ using \eqref{cf}  \\
$\mu=1$,  $\lambda=1$  & $10$ & $100$ & $0.870$ & NA & NA\\\hline
\end{tabular}
\caption{Summary of simulations.}\label{tbSimRes}
\end{center}
\end{table}

In the critical case the correlation coefficient $\rho_n$ is undefined as both the 
covariance between two sampled species and 
the species' variances take infinite values. We overcome this difficulty by modifying the Aldous--Popovic approach, 
we replace the improper prior distribution for $T$ by the uniform prior on a finite interval $(0,N)$. We believe 
that considering a proper prior in the critical case makes the model biologically more relevant. 
A realistic value of $N$ gives an upper bound on the number of speciation events for a group of 
related species as traced back to their common ancestor. This number depends on the particular 
kind of organisms in consideration and in many cases cannot be larger than several thousands.

In Section \ref{sec:VE}
the obtained formulae for $\rho_n$ and $\E{T}$ are combined with Eqs. \eqref{eqVarXnBM} and \eqref{eqES2BM} to produce compact expressions for $\var{\bar{X}_{n}}$ and $\E{S^{2}_{n}} $ in the three main cases. These analytic expressions can be used, for example,
to construct phylogenetic confidence intervals for the ancestral trait value $X_0$, which would take into account 
tree uncertainty. This issue is one of the subjects of our forthcoming paper \
where among other things some of the results of this paper
for the Brownian motion model are extended to the Ornstein--Uhlenbeck model.

Section \ref{scoi} presents a connection to a new measure of how balanced are phylogenetic trees  recently introduced by \citet{2012arXiv1202.1223M}. 
\ref{appA} and \ref{appB} contain intermediate results. \ref{appB} is mainly dealing with the properties of an important for this paper expression,
\be
e_{n,m}=m^{n}\left(\ln {m\over m-1}-\sum\limits_{i=1}^{n}{1\over im^{i}}\right)=\sum\limits_{i=1}^{\infty}{1\over (n+i)m^{i}},
\label{enl}
\ee
which satisfies $0<e_{n,m}<{1\over n (m-1)}$ for $1<m<\infty$. 

\section{Correlation coefficient for the Yule model}\label{sy}
Assume that the trait values evolve according to Brownian motions with variance $\sigma^2$. 
At the time of origin the ancestral trait is believed to have a fixed value $X_0$. Due to the formula for
the total variance, the variance of a sampled trait value equals,
\bd
\begin{array}{rcl}
\var{X_i}&=&\E{\var{X_i|T}}+\Var{\E{X_i|T}}\\
&=&\E{\sigma^2T}+\Var{0}=\sigma^2\E{T}.
\end{array}
\ed
If $X_{ij}$ is the ancestral trait value at the time $\tau_{ij}$ of the most recent common ancestor for two sampled species, then
\bd
\begin{array}{rcl}
\Cov{X_i,X_j}&=&\E{\Cov{X_i,X_j|T,\tau_{ij},X_{ij}}} +\Cov{\E{X_i|T,\tau_{ij},X_{ij}},\E{X_j|T,\tau_{ij},X_{ij}}}\\
&=&\E{0}+\Var{X_{ij}}=\sigma^2\E{T-\tau_{ij}}.
\end{array}
\ed
Putting this into \eqref{defrho} we get
\be\label{ro}
\rho_n={\E{T-\tau}\over\E{T}},
\ee
where $\tau$ is the time to the most recent common ancestor for a pair of randomly chosen extant species.

The denominator in Eq. \eqref{ro} is computed as
\bd
\E{T}=\int\limits_0^\infty tq_n(t)\ud t,
\ed
where $q_n(t)$ is the  distribution density for the time to origin $T$. 
Assuming the Yule model with rate $\lambda=1$ for the unknown species tree it is easy to see that
\be\label{yt}
\E{T}=a_n.
\ee
Indeed, using the formula \citep[see][]{TGer2008}
\be
q_n(t)=n(1-e^{-t})^{n-1}e^{-t},\label{qnY}
\ee
and applying a change of variables
\bd
v=1-e^{-t},\ t=-\ln (1-v),\ \ud v=(1-v)\ud t,
\ed
we conclude
\bd
\begin{array}{rcl}
\E{T}&=&-n\int\limits_0^1\ln (1-v)v^{n-1}dv=-\int\limits_0^1\ln (1-v)\ud (v^{n}-1)\\
&=&\int\limits_0^1{1-v^n\over1-v}\ud v=a_n.
\end{array}
\ed

In the framework of the conditioned reconstructed process model \citep[see][]{TGer2008}
the random species tree (extinct species removed) is conveniently described in terms of speciation times $s_1,\ldots,s_{n-1}$, 
see panel C in Fig. \ref{tr}. Conditioned on the time of origin $T=t$ the random variables $s_1,\ldots,s_{n-1}$ are independent and 
identically distributed according to a cumulative distribution function to be denoted by $F_t(s)$. Due to this observation
the numerator Eq. \eqref{ro} can be found from the formula
\be
\E{T-\tau}=\sum\limits_{k=1}^{n-1}{2(n-k)\over n(n-1)}\int\limits_0^\infty\left(\int\limits_0^t F^k_t(s)ds\right)q_n(t)\ud t,
\label{tmt}
\ee
derived in \ref{appA}.

\begin{figure}
\begin{center}
\includegraphics[width=0.32\textwidth]{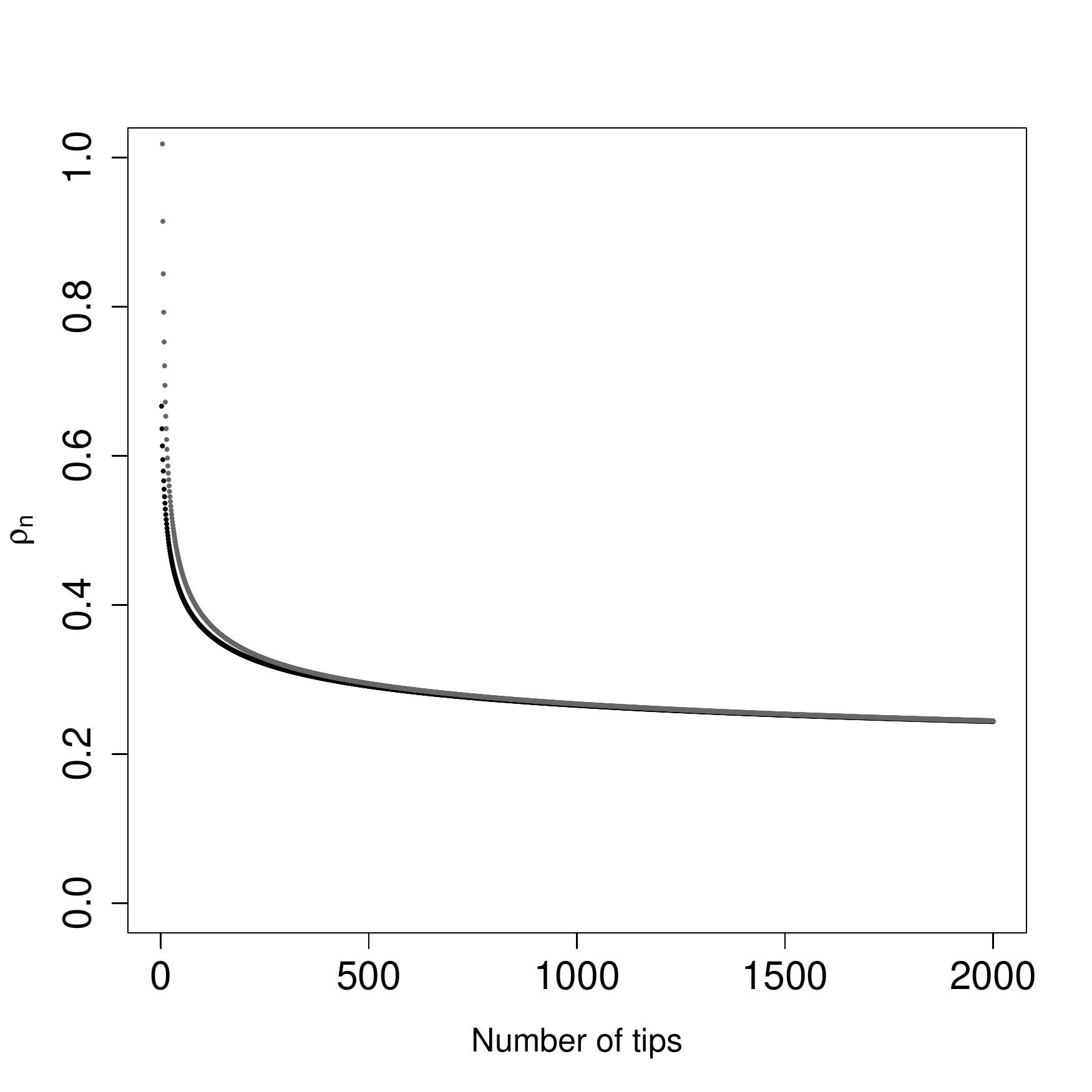}
\includegraphics[width=0.32\textwidth]{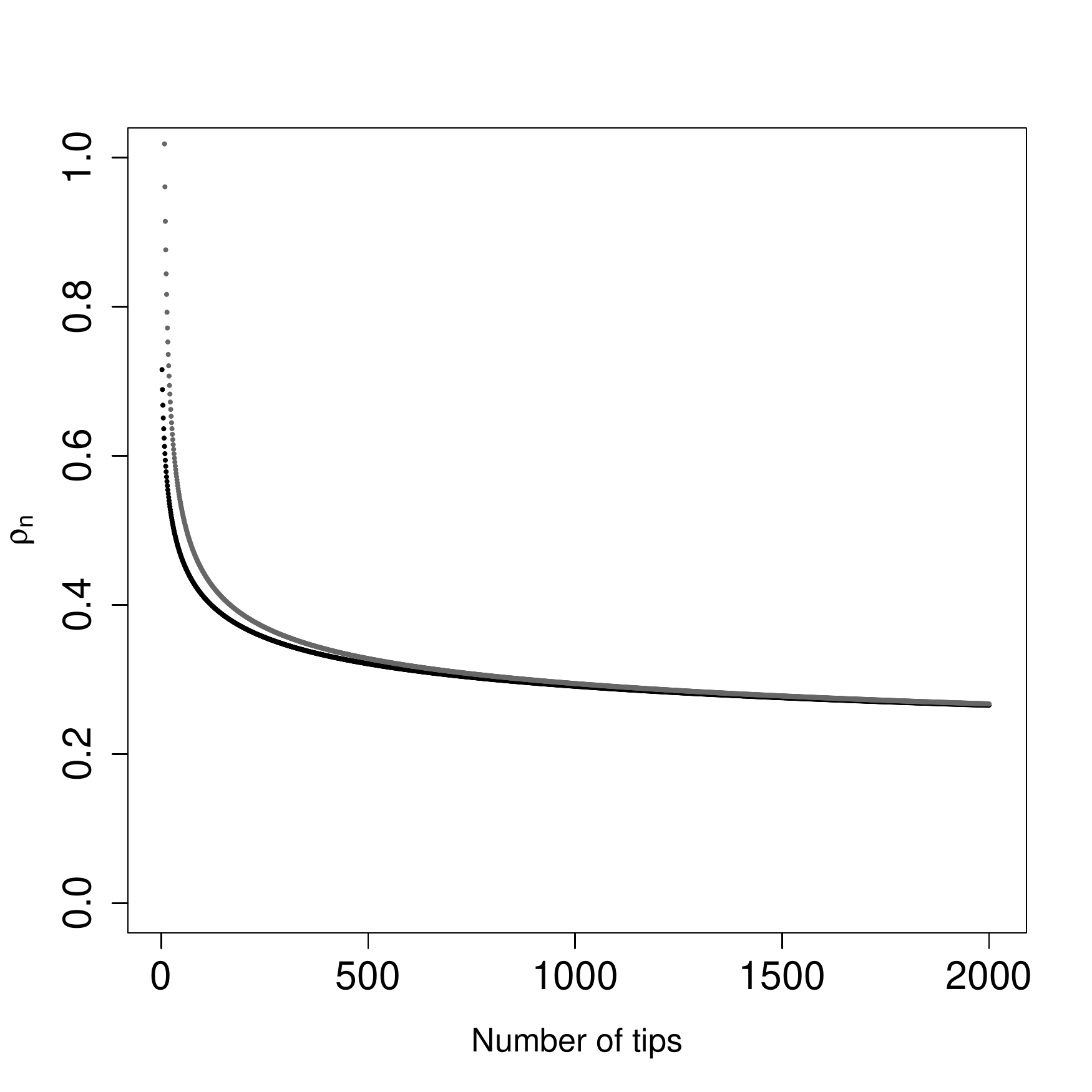}
\includegraphics[width=0.32\textwidth]{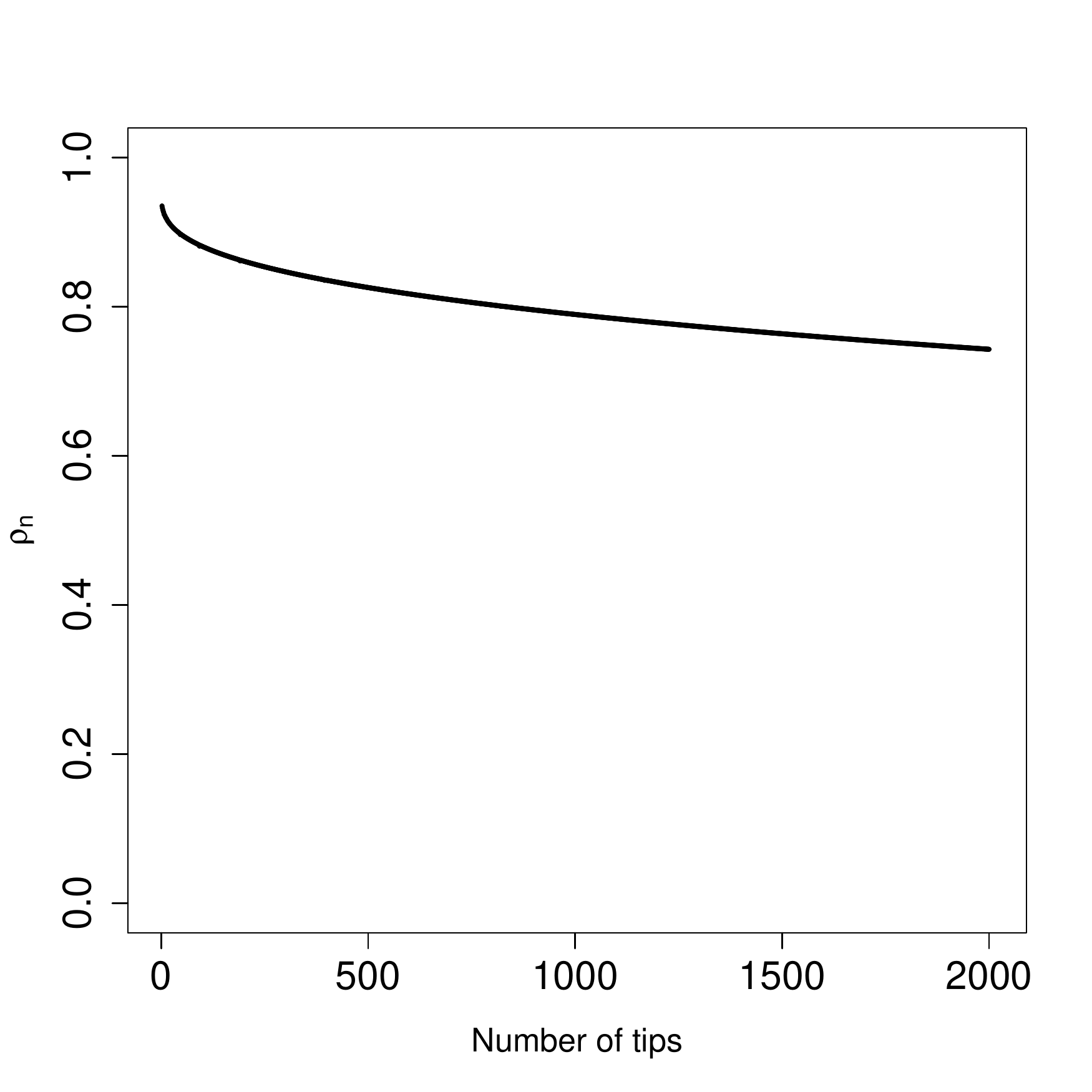}
\end{center}
\caption{Exact (black line) and approximate (gray line) formulae for $\rho_n$ in the 
Yule case (left), supercritical case $\lambda=2$ (centre) 
and critical case with proper prior $N=10000$ (right).}
\label{fY}
\end{figure}

In the Yule case we have
\be\label{F1}
F_t(s)={1-e^{-s}\over1-e^{-t}}1_{\{0<s\le t\}}+1_{\{s> t\}}
\ee
which together with Eq. \eqref{tmt} after applying a change of variables $u=1-e^{-s}, v=1-e^{-t}$ gives
\bd
\begin{array}{rcl}
\E{T-\tau}&=&{2\over n-1}\sum\limits_{k=1}^{n-1}(n-k)\int\limits_0^1\int\limits_0^v {u^kv^{n-k-1}\over1-u}\ud u\ud v\\
&=&{2\over n-1}\sum\limits_{k=1}^{n-1}\int\limits_0^1\int\limits_u^1 {u^k\over1-u}\ud v^{n-k}\ud u.
\end{array}
\ed
Switching the integration order we find,
\bd
\begin{array}{rcl}
\E{T-\tau}&=&{2\over n-1}\sum\limits_{k=1}^{n-1}\int\limits_0^1{(1-u^{n-k})u^k\over1-u}\ud u\\
&=&{2\over n-1}\sum\limits_{k=1}^{n-1}(a_n-a_k)={2(n-a_n)\over n-1}.
\end{array}
\ed
Combining this with Eqs. \eqref{ro} and  \eqref{yt} we arrive at 
\be\label{yf}
\rho_n={2\over n-1}\left({n\over a_n}-1\right),
\ee
where $a_n=\sum_{i=1}^{n}{1\over i}$ is the $n$-th harmonic number. Notice that Eq. \eqref{yf} implies,
\be\label{ln}
\rho_n={2\over\ln n+\gamma+o(1)},\ n\to \infty,
\ee
where $\gamma=0.577\ldots$ is the Euler constant, in other words, ${2\over\rho_n}-\ln n-\gamma\to0$ as $n\to \infty$.
The exact formula Eq. \eqref{yf} and the approximate formula Eq. \eqref{ln}, with the term $o(1)$ being disregarded, are illustrated in Fig. \ref{fY}, left panel.

\section{Supercritical case}\label{ss}
In the supercritical case the correlation coefficient has a more complicated but still surprisingly compact form in terms 
of the function from Eq. \eqref{enl},
\be\label{sf}
\rho_n={2\over n-1}\left({n(1+e_{n,\lambda})\over  a_n+e_{n,\lambda}-\ln{\lambda\over \lambda-1}}-{\lambda\over \lambda-1}\right).
\ee
Observe that Eq. \eqref{yf} can be recovered from Eq. \eqref{sf} by letting $\lambda\to\infty$.
Furthermore, Eq. \eqref{sf} implies a close counterpart of Eq. \eqref{ln},
\be\label{ln1}
\rho_n={2\over\ln n+\gamma-\ln{\lambda\over \lambda-1}+o(1)},\ n\to \infty
\ee
uniformly in $\lambda\ge\lambda_0$ for any $\lambda_0>1$. 

\begin{figure}
\begin{center}
\includegraphics[width=0.32\textwidth]{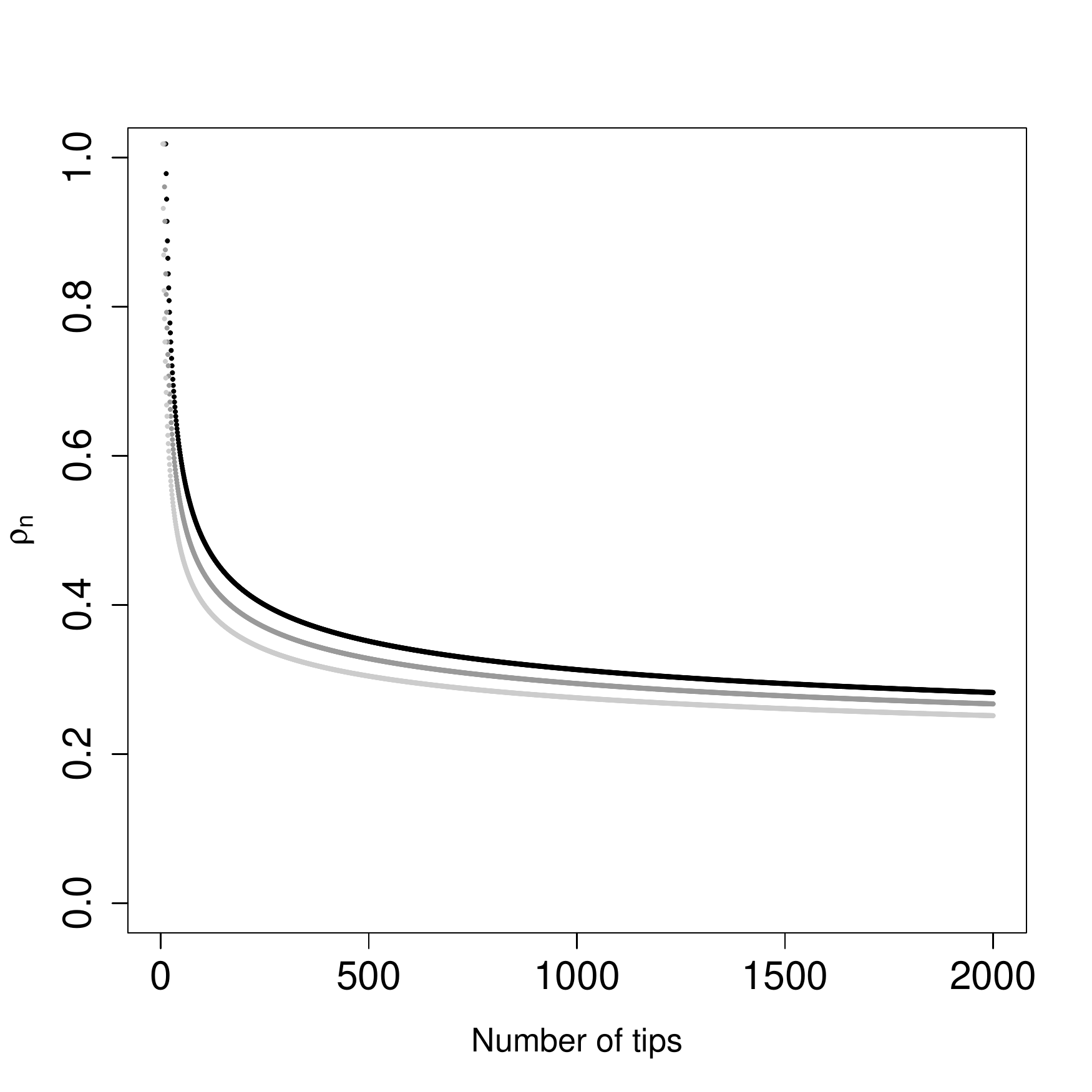}
\includegraphics[width=0.32\textwidth]{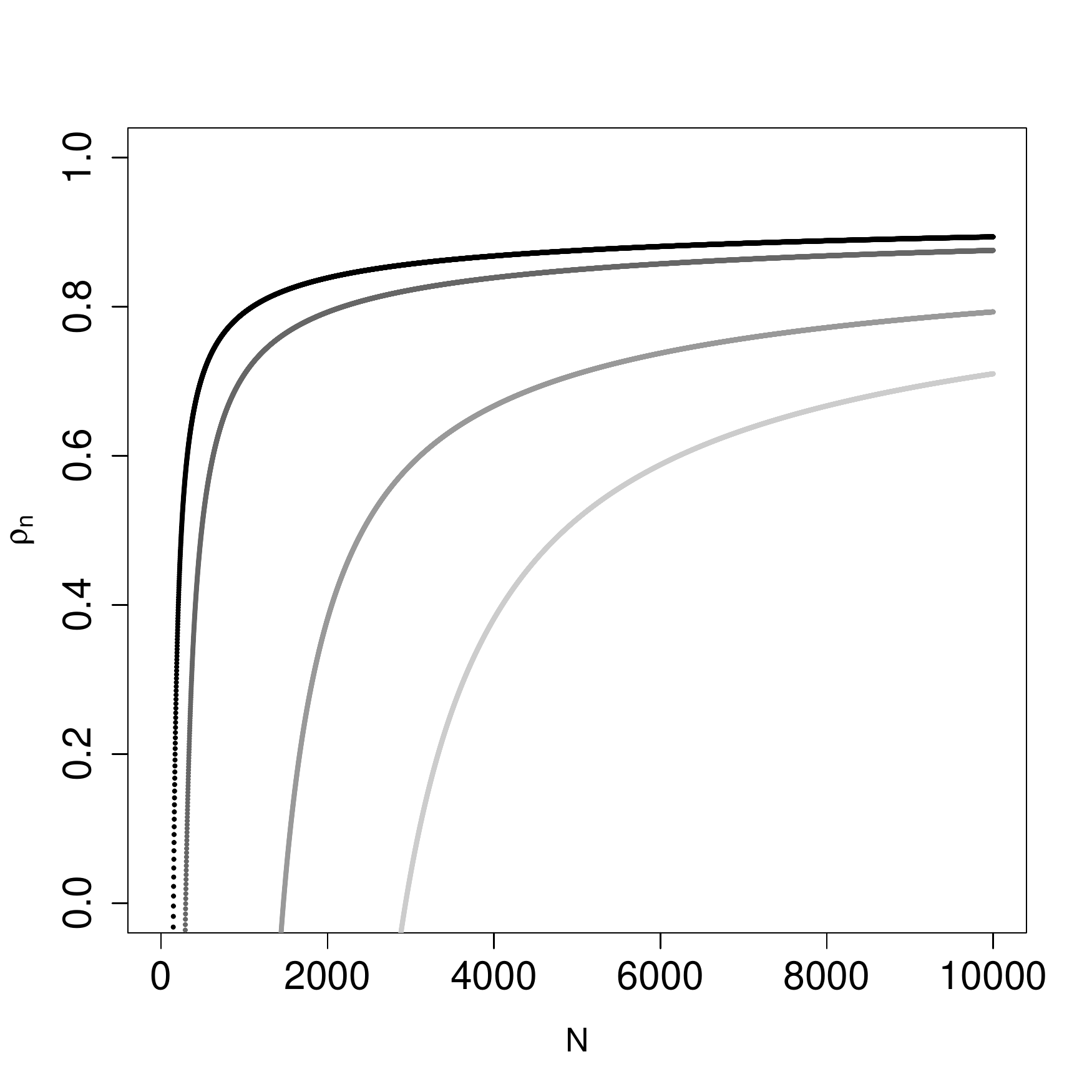} 
\includegraphics[width=0.32\textwidth]{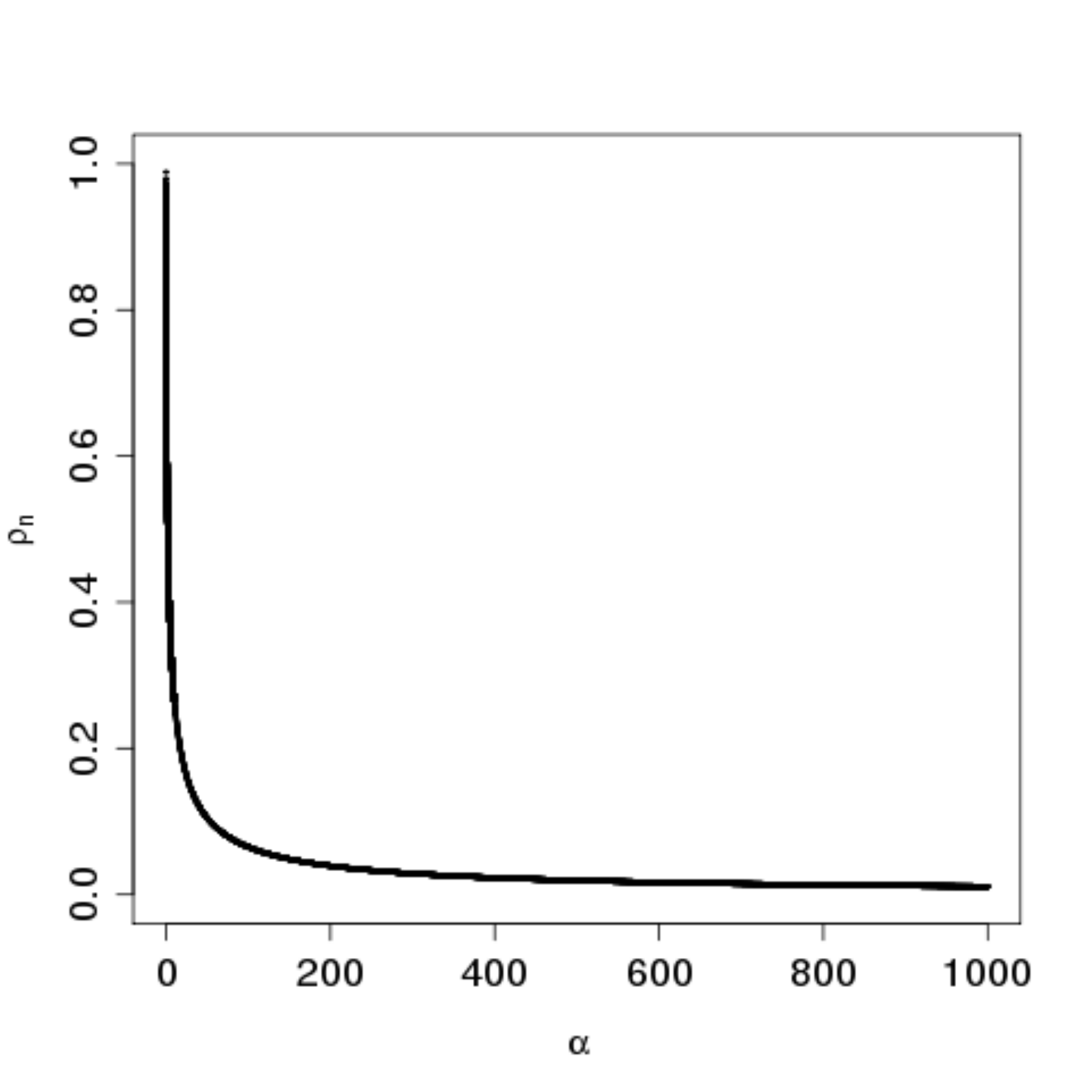} 
\end{center}
\caption{Approximated correlation coefficient as a function of model parameters. 
Left, the supercritical case, Eq. \eqref{ln1}: black $\lambda = 1.5$, gray $\lambda = 2$, light gray $\lambda = 5$. 
Center, the critical case with a proper prior, Eq. \eqref{cf2}: black $n=50$, gray $n=100$, light gray $n= 500$, lightest gray $n = 1000$.
Right: the critical case with a proper prior, Eq. \eqref{cfa1}.
}
\label{fsc}
\end{figure}

Specializing on the nearly critical case, when $\lambda=1+1/N$ for some large $N$, we derive the following asymptotic result,
\be\label{cf}
\rho_n= 1-{1\over2(\ln N-a_n+1)+o(1)},\ N\to\infty.
\ee
The  fact that $\rho_n\to1$ as $N\to\infty$ is a consequence of the 
improper prior distribution
assumption for the time of 
origin $T$. Besides this approximation, it can be shown that for any fixed positive $\alpha$,
\be\label{cfa}
\rho_n\to 2\left({1+I_\alpha\over \ln \alpha+\gamma+I_\alpha}-{1\over\alpha}\right),\ N\to\infty,\ n/N\to\alpha,
\ee
where $I_\alpha=\int_0^\infty {e^{-x}\ud x\over\alpha+x}$, so that $e^{-\alpha}I_\alpha=\int_\alpha^\infty {e^{-x}\ud x\over x}$ is the exponential integral.
The exact formula Eq. \eqref{sf} and the approximate formula Eq. \eqref{ln1} are illustrated in Fig. \ref{fY}, central panel. Another illustration of Eq. \eqref{ln1} is given on Fig. \ref{fsc}, left panel. 

We derive Eq. \eqref{sf} using,
\bd
\begin{array}{rcl}
q_n(t)&=&n\lambda^{n}(\lambda-1)^2{x_t^{n-1}(1-x_t)\over(\lambda-1+x_t)^{n+1}},\\
F_t(s)&=&{x_s\over \lambda-1+x_s}{\lambda-1+x_t\over x_t},
\end{array}
\ed
where $x_t=1-e^{-(\lambda-1)t}$. These expressions are obtained from more general relations due to \citet{TGer2008} after specifying 
the parameter values as $\mu=1$ and $\lambda>1$. Denoting $\delta=\lambda^{-1}$ we can write,
\bd
\E{T}=n\delta(\lambda-1)^2\int\limits_0^\infty {x_t^{n-1}(1-x_t)\over( 1-\delta(1-x_t))^{n+1}}t\ud t.
\ed
A change of variables,
\bd
\begin{array}{ll}
v={x_t\over1-\delta(1-x_t)},&1-x_t={1-v\over 1-\delta v},\\
(\lambda-1)t=\ln {1-\delta v\over 1-v},&\ud v=\lambda(1-\delta v)(1-v)\ud t,
\end{array}
\ed
results in,
\bd
\begin{array}{rcl}
\E{T}&=&n(\lambda-1)^{-1}\int\limits_0^1\left(\ln {1-\delta v\over 1-v}\right)v^{n-1}\ud v\\
&=&(\lambda-1)^{-1}\int\limits_0^1\left(\ln (1-\delta v)-\ln (1-v)\right)\ud (v^{n}-1)\\
&=&\delta\int\limits_0^1{1-v^{n}\over(1-\delta v)(1-v)}\ud v=\sum\limits_{k=0}^{n-1}\int\limits_0^1{v^k\over\lambda- v}\ud v.
\end{array}
\ed
Applying Eqs. \eqref{e3} and \eqref{e1} from \ref{appB}, leads to,
\be
\E{T}={1\over \lambda-1}\left(a_n+e_{n,\lambda}-\ln{\lambda\over \lambda-1}\right).\label{st}
\ee

Furthermore, with $u={x_s\over1-\delta(1-x_s)}$ and $\ud u=\lambda(1-\delta u)(1-u)\ud s$ Eq. \eqref{tmt} entails,
\bd
\begin{array}{rcl}
\E{T-\tau}&=&{2\over n-1}\sum\limits_{k=1}^{n-1}(n-k)\int\limits_0^1\int\limits_0^v {u^kv^{n-k-1}\over(\lambda- u)(1-u)}\ud u\ud v\\
&=&{2\over \lambda(n-1)}\sum\limits_{k=1}^{n-1}\int\limits_0^1\int\limits_u^1 {u^k\over(1-\delta u)(1-u)}\ud v^{n-k}\ud u.
\end{array}
\ed
Due to,
\bd
\sum\limits_{k=1}^{n-1}\int\limits_0^1\int\limits_u^1 {u^k\over(1-\delta u)(1-u)}\ud v^{n-k}\ud u=\sum\limits_{k=1}^{n-1}\sum\limits_{i=0}^{n-k-1}\int\limits_0^1{u^{k+i}\ud u\over1-\delta u}=
\sum\limits_{j=1}^{n-1}\int\limits_0^1{ju^j\ud u\over1-\delta u}
\ed
we arrive at,
\bd
\begin{array}{rcl}
\E{T-\tau}&\stackrel{\eqref{e3}}{=}&{2\over n-1}\sum\limits_{k=1}^{n-1}ke_{k,\lambda}\\
&\stackrel{\eqref{e2}}{=}&{2\over (n-1)(\lambda-1)}\left(n+ne_{n,\lambda}-{\lambda\over \lambda-1}
\left(a_n+e_{n,\lambda}-\ln{\lambda\over \lambda-1}\right)\right).
\end{array}
\ed
which together with Eqs. \eqref{st} and \eqref{ro} gives Eq. \eqref{sf}.

\section{Critical case with a proper prior}\label{sc}
Under the improper uniform prior on $(0,\infty)$ for $T$ one has \citep[see][]{DAldLPop},
\bd q_n(t)={nt^{n-1}\over(1+t)^{n+1}},\ t\in(0,\infty)\ed
implying the infinite mean $\E{T}$. To remedy this inconvenience we use a proper uniform prior on $(0,N)$ and put $m={N+1\over N}$. 
The corresponding posterior distribution of $T$ has density,
\bd q_n(t)={nt^{n-1}m^{n}\over(1+t)^{n+1}},\ t\in(0,N)\ed
with finite mean,
\be
\E{T}=ne_{n,m},\label{ct}
\ee
obtained as
\bd
\E{T}=nm^{n}\int\limits_0^N {t^n \ud t\over(1+t)^{n+1}}=nm^{n}\int\limits_0^{1/m} {x^n \ud x\over1-x}\stackrel{\eqref{e3}}{=}ne_{n,m}.
\ed

For the critical case with a proper prior we establish
\be\label{cf1}
\rho_n=2-{2N\over n-1}\left(1+{1\over e_{n,m}}\right)+{2N(N+1)\over n(n-1)}\left(1+{a_n-\ln(N+1)\over e_{n,m}}\right),
\ee
where $m=1+1/N$. Interestingly, the following approximate version of Eq. \eqref{cf1},
\be\label{cf2}
\rho_n= 1-{1\over2(\ln N-a_n)+o(1)},\ N\to\infty.
\ee
is almost the same as Eq. \eqref{cf}. The counterpart of Eq. \eqref{cfa} is given by,
\be\label{cfa1}
\rho_n\to 2-{2\over\alpha}\left(1+{1\over I_\alpha}\right)+{2\over\alpha^2}\left(1+{\ln\alpha+\gamma\over I_\alpha}\right),\ N\to\infty,\ n/N\to\alpha.
\ee
Eq. \eqref{cf1} is illustrated on the right panel of Fig. \ref{fY}, while Eqs. \eqref{cf2} and \eqref{cfa1} are illustrated on the central and right panels of Fig. \ref{fsc}.

Next we derive Eq. \eqref{cf1} using the formula $F_t(s)={s(1+t)\over(1+s) t}$ obtained by \citet{DAldLPop}. Entering this into Eq. \eqref{tmt} gives,
\bd
\E{T-\tau}={2m^{n}\over n-1}\sum\limits_{k=1}^{n-1}(n-k)\int\limits_0^N \int\limits_0^t \left({s\over1+s}\right)^k\left({1+t\over t}\right)^{k-n+1}{1\over(1+t)^{2}}\ud s\ud t.
\ed
Replacing the variables $s$ and $t$  with $y={s\over1+s}$ and $x={t\over1+t}$ we get,
\bd
\begin{array}{rcl}
\E{T-\tau}&=&{2m^{n}\over n-1}\sum\limits_{k=1}^{n-1}\int\limits_0^{1/m}\int\limits_0^x y^k(1-y)^{-2}\ud y\ud x^{n-k}\\
&=&{2\over n-1}\sum\limits_{k=1}^{n-1}\int\limits_0^{1/m}(my)^k(1-y)^{-2}(1-(my)^{n-k})\ud y
\end{array}
\ed
and then,
\bd
\begin{array}{rcl}
\E{T-\tau}&\stackrel{\eqref{e4}}{=}&{2\over n-1}\sum\limits_{k=1}^{n-1}\left(ne_{n,m}-ke_{k,m}\right)\\
&\stackrel{\eqref{e2}}{=}&2ne_{n,m}-{2\over n-1}\left({ne_{n,m}+n\over m-1}-{m\over (m-1)^2}\left(a_n+e_{n,m}-\ln{m\over m-1}\right)\right),
\end{array}
\ed
which combined with Eq. \eqref{ct} gives Eq. \eqref{cf1}.

\section{Variance of sample mean and expectation of sample variance}\label{sec:VE}
Our formulae for $\rho_n$ and $\E{T}$ obtained in the previous sections imply the following compact expressions for $\var{\bar{X}_{n}}$ and $\E{S^{2}_{n}} $ thanks to 
Eqs. \eqref{eqVarXnBM} and \eqref{eqES2BM}.

\begin{figure}
\begin{center}
\includegraphics[width=0.50\textwidth]{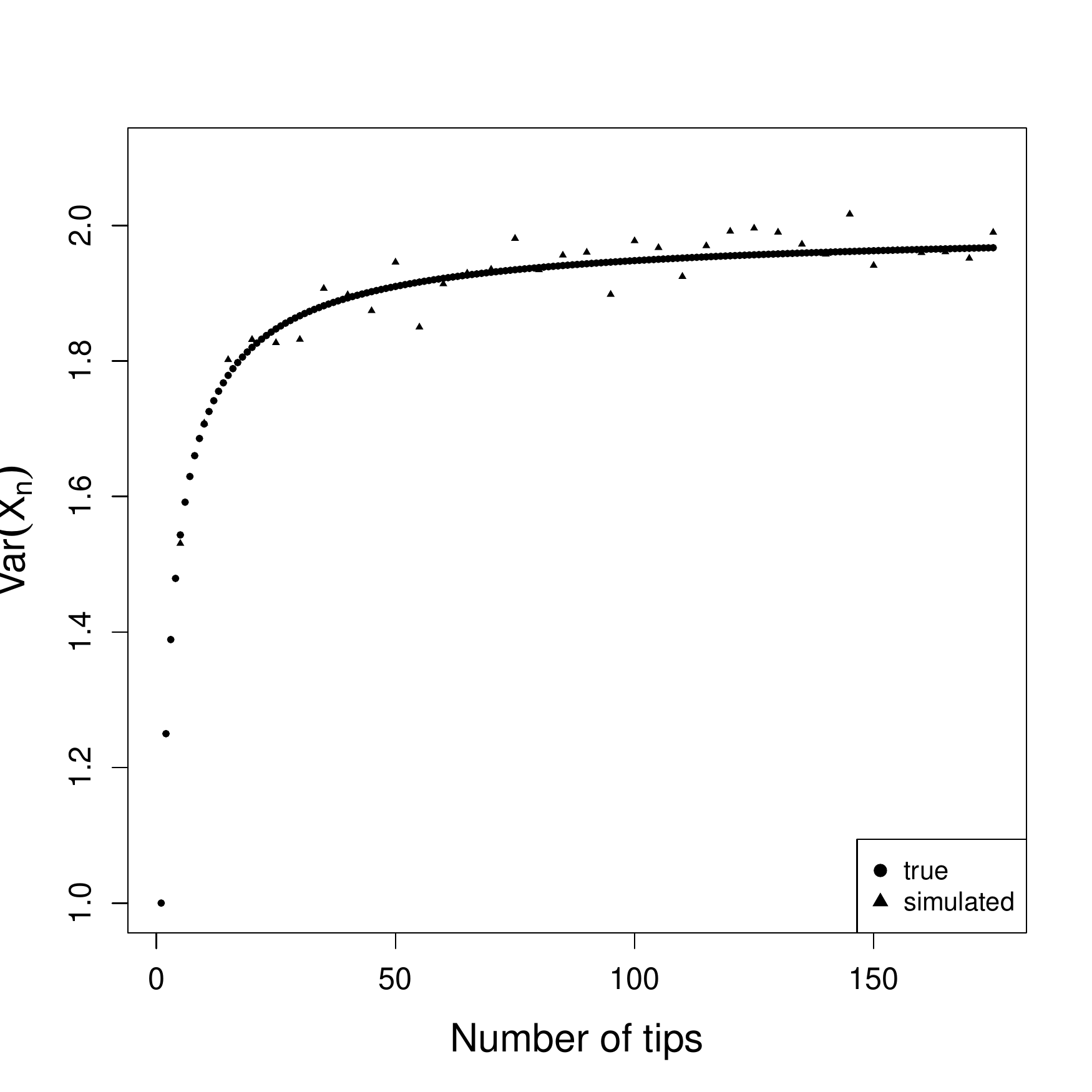}
\end{center}
\caption{Variance of sample mean in the Yule case given by Eq. \eqref{eqVarXnBMYule} with points indicating simulated values. Each point
is the estimate of the variance based on $10000$ simulations for each value of $n$.
In simulations $X_{0}=0$ and $\sigma^{2}=1$.
}
\label{figYuleBMVarXn}
\end{figure}

In the Yule case ($\mu=0$ and  $\lambda=1$)  Eqs. \eqref{yt} and \eqref{yf}  give
\be\label{eqVarXnBMYule}
\var{\bar{X}_{n}} = \sigma^{2}(2-\frac{a_{n}}{n}).
\ee
In Fig. \ref{figYuleBMVarXn} we can see that the above formula and its consequence $\var{\bar{X}_{n}}  \rightarrow 2\sigma^{2}~\mathrm{as}~n\rightarrow \infty$ agree well with simulations.
Notice that this immediately implies that the unbiased point estimate $\bar X$ of the ancestral state $X_{0}$ is
not consistent as the variance of the estimator tends to a constant $2\sigma^{2}$. 
We can compare this with the result of \citet{CAne} who deals with another estimator of the ancestral state. 
The estimator of \citet{CAne} is unbiased and converges (in $L^{2}$ and almost surely) to a random variable with a non--zero variance, bounded from below by
$\sigma^{2}t/k$, where $t$ is the maximum length of a branch stemming from the root
and $k$ is the number of branches stemming from the root ($k=2$ in our model). 
However, \citet{CAne} considers a different model of tree growth as $n\rightarrow \infty$ 
and the tree is assumed to start at the root.

Using Eqs. \eqref{eqES2BM} and \eqref{eqVarXnBMYule} we obtain for the Yule case
\be
\E{S^{2}_{n}} = \sigma^{2}\left(\frac{n+1}{n-1}a_{n} -\frac{n}{n-1}2\right),
\ee
so that $\E{S^{2}_{n}} \sim\sigma^{2}\ln n$ as $n\to\infty$. This suggest an unbiased estimate for the variance $\sigma^{2}$.

In the supercritical case ($\mu=1$ and $\lambda>1$) Eqs. \eqref{st} and \eqref{sf} entail
\bd
\begin{array}{rcl}
\var{\bar{X}_{n}} &= &2\sigma^{2}\left(\frac{1+e_{n,\lambda}}{\lambda-1}-\frac{1}{n}\frac{\lambda+1}{(\lambda-1)^{2}}\left(a_{n}+e_{n,\lambda}-\ln \frac{\lambda}{\lambda-1}\right) \right),\\
\E{S^{2}_{n}} &= &\sigma^{2} \frac{\left(\frac{2\lambda}{\lambda-1}+n\right)(a_{n}-\ln  \frac{\lambda}{\lambda-1}) +(\frac{2\lambda}{\lambda-1}-n)e_{n,\lambda}+1-2n}{(\lambda -1)(n-1)}.
\end{array}
\ed
In the critical case ($\mu=\lambda=1$) with a proper prior imposed on the
time of origin we use  Eqs. \eqref{ct} and \eqref{cf1} to get
\bd
\begin{array}{rcl}
\var{\bar{X}_{n}} &=\sigma^{2}\left( e_{n,m}\left(2n-2N+{2N(N+1)\over n}-1\right)- \frac{A_{n,N}}{n}\right),\\
\E{S^{2}_{n}}& =\sigma^{2}\left(e_{n,m}\left(\frac{2Nn}{n-1}-{2N(N+1)\over n-1}-n\right)+\frac{A_{n,N}}{n-1}\right),
\end{array}
\ed
where $m=1+1/N$ and $$A_{n,N}=2N(n-(N+1)(a_{n}-\ln(N+1))).$$

To compare different cases we put together asymptotic formulae as $n\to\infty$ (and additionally $n/N\to\alpha$ in the critical case):
\be\label{sumVarXn}
\sigma^{-2}\var{\bar{X}_{n}} \sim\left\{ 
\begin{array}{ll}
2&\mbox{ in the Yule case,}\\
{2\over\lambda-1}&\mbox{ in the supercritical case,}\\
c_\alpha n&\mbox{ in the critical case},
\end{array}
\right.
\ee 
where $c_\alpha=2\alpha^{-2}((\alpha^2-\alpha+1)I_\alpha-\alpha+\ln \alpha)$, and
\be\label{sumES2}
\sigma^{-2}\E{S^{2}_{n}}\sim\left\{ 
\begin{array}{ll}
\ln n&\mbox{ in the Yule case,}\\
{\ln n\over\lambda-1}&\mbox{ in the supercritical case,}\\
d_\alpha n&\mbox{ in the critical case},
\end{array}
\right.
\ee 
where $d_\alpha=\alpha^{-2}((2\alpha-\alpha^2-2)I_\alpha+\alpha-\ln \alpha)$.

\section{Connection with total cophenetic index}\label{scoi}
A recent work due to \citet{2012arXiv1202.1223M} considers a new balance index for phylogenetic trees termed the \textit{total cophenetic index}. The total cophenetic index for a given tree with $n$ tips is defined
as, 
\bd
\Phi_n= \sum\limits_{1 \le i < j \le n}\phi_{ij},
\ed
the sum of the number of branches $\phi_{ij}$ from the root to the most recent common
ancestor of tips $i$ and $j$. Their model of the phylogenetic tree is different from the one we discuss in that there is no branch prior to the
root, i.e. the tree ``begins'' at the first branching point. Under the Yule model they show that the 
expectation of the total cophenetic index for a tree with $n$ tips is 
\be
\E{\Phi_{n}}=n(n+1-2a_n).
\label{coi}
\ee

We next demonstrate a short proof of the latter formula based on the approach developed in this paper. 
Denote by $T_n$ the time of the tree root so that $T-T_n$ is the length of the initial branch until the first splitting 
(for illustration see Fig. \ref{tr}, panel A). For the conditional Yule tree with $n$ tips the random variable
\bd
\Phi^{\ast}_{n} =  \sum\limits_{1 \le i < j \le n}(T_n-\tau_{ij})
\ed
is the sum of branch lengths connecting the root with the
most recent common ancestor of tips $i$ and $j$. Since the mean branch length of this random tree is 
$0.5$ \citep[see][for results on branch length expectations]{AMooetal,TStaMSte2012}  we have $\E{\Phi_{n}}=2\E{\Phi^{\ast}_n}$, 
and Eq. \eqref{coi} follows from
\bd
\E{\Phi^{\ast}_n}  = \E{\sum\limits_{1\le i < j \le n}(T_n-\tau_{ij})} = \binom{n}{2}\E{T_n-\tau},
\ed
where as before $\tau$ is the time to the most recent common ancestor for a randomly chosen pair of tips. Indeed, using
the simple fact proved in \ref{appA},
\be
\E{T-T_n}=1,\label{Tn}
\ee
we get the required equality
\bd
\begin{array}{rcl}
\E{\Phi^{\ast}_n} & = &\binom{n}{2}(\E{T-\tau}-1)\\
&=&\binom{n}{2}\left(\frac{2(n-a_{n})}{n-1}-1\right)
=\frac{n}{2}(n+1-2a_{n}).
\end{array}
\ed

\section*{Acknowledgments}
The research of Serik Sagitov was supported by the Swedish Research Council grant 621-2010-5623. 
The research of Krzysztof Bartoszek was partially supported by the Center of Theoretical Biology at the University of Gothenburg.
We would like to thank Graham Jones for numerical procedures for calculation of $e_{n,m}$ and providing R code for this.

\appendix
\section{}\label{appA}
This section contains derivation of formulae \eqref{eqVarXnBM}, \eqref{eqES2BM},  \eqref{tmt}, and \eqref{Tn}.

Relations \eqref{eqVarXnBM} and \eqref{eqES2BM}  come straightforwardly from 
\bd
\begin{array}{rcl}
n^{2}\var{\bar{X}_{n}} &= &\var{\sum\limits_{i=1}^{n}X_{i}}\\
&=&n\var{X} +  2\sum\limits_{1\le i< j\le n}\Cov{X_{i},X_{j}}\\
&=&\frac{1}{n}(1+(n-1)\rho_{n})\var{X}
\end{array}
\ed
and
\bd
\begin{array}{rcl}
\E{S^{2}_{n}} & = & \frac{n}{n-1}\E{\frac{1}{n}\sum\limits_{i=1}^{n}X_{i}^{2} -(\bar{X}_{n})^{2}}
\\ & = & 
\frac{n}{n-1}(\E{X^{2}}-\E{\bar{X}_{n}}^{2}) 
= 
\frac{n}{n-1}(\var{X} - \var{\bar{X}_{n}}) 
\\ & = &(1-\rho_{n})\var{X}
\end{array}
\ed
due to the variance formula $\var{X}=\sigma^{2}\E{T}$ characterizing the Brownian motion model of evolution considered here.

Equation \eqref{tmt} is obtained as follows. If $\kappa$ is the corresponding distance between the sampled tips, then
\be
\mathbb P(\kappa=k)={n-k\over {n\choose2}},\ k=1,\ldots,n-1,
\ee 
because in a row of $n$ positions there are $n-k$ pairs on a distance $k$. Since $\tau$ is the maximum of $\kappa$ 
independent and identically distributed speciation times, we get,
\bd \mathbb P(\tau>s|T,\kappa)=1-F^\kappa_T(s),\ed
and Eq. \eqref{tmt} follows from,
\bd \E{T-\tau|T,\kappa}=T-\int\limits_0^T (1-F^\kappa_T(s))ds=\int\limits_0^T F^\kappa_T(s)\ud s.\ed
Finally,  \eqref{Tn} follows from  \eqref{qnY} and \eqref{F1}
\bd
\begin{array}{rcl}
\E{T-T_n}&=&\E{\int\limits_0^T F^{n-1}_T(s)ds}\\
&=&\int\limits_0^\infty ne^{-t}\int\limits_0^t (1-e^{-s})^{n-1}\ud s\ud t\\
&=&n\int\limits_0^1\int\limits_u^1(1-v)^{n-1}v^{-1}\ud v\ud u\\
&=&n\int\limits_0^1(1-v)^{n-1}\ud v=1.
\end{array}
\ed

\section{}\label{appB}
In the main text we use the following relations for the function in Eq. \eqref{enl}:
\be
\begin{array}{rcl}
e_{k,m}&=&\int\limits_0^1{x^k\ud x\over m-x},\label{e3}\\
\end{array}
\ee
\be
\begin{array}{rcl}
\sum\limits_{k=0}^{n-1}e_{k,m}&=&{1\over m-1}\left(a_n+e_{n,m}-\ln{m\over m-1}\right),\label{e1}\\
\end{array}
\ee
\be
\begin{array}{rcl}
\sum\limits_{k=1}^{n-1}ke_{k,m}&=&{n+ne_{n,m}\over m-1}-{m\left(a_n+e_{n,m}-\ln{m\over m-1}\right)\over (m-1)^2},\label{e2}\\
\end{array}
\ee
\be
\begin{array}{rcl}
m^{k}\int\limits_0^{1/m} y^k(1-y)^{-2}\ud y&=&{1\over m-1}-ke_{k,m}\label{e4}.
\end{array}
\ee
Equation \eqref{e3} follows from,
\bd
\begin{array}{rcl}
\int\limits_0^1{v^k\over1-vm^{-1}}\ud v&= &m^{k+1}\int\limits_0^{1/m} {x^k\ud x\over1-x}=m^{k+1}\int\limits_0^{1/m} \left({1\over1-x}-{1-x^k\over1-x}\right)\ud x\\
&=&m^{k+1}\ln{m\over m-1}-\sum\limits_{i=1}^{k}{m^{k+1-i}\over i}=m e_{k,m}.
\end{array}
\ed
To prove Eq. \eqref{e1} put $x=\sum_{k=0}^{n-1}e_{k,m}$ and using,
\be
me_{k-1,m}=k^{-1}+e_{k,m},\label{ek}
\ee
set up a linear equation,
$$mx=a_n+x +e_{n,m}-\ln{m\over m-1}$$
whose solution is Eq. \eqref{e1}.
Similarly, to obtain Eq. \eqref{e2} put $x=\sum_{k=1}^{n-1}ke_{k,m}$ and use Eq. \eqref{ek} to get a linear equation, 
\bd \sum\limits_{k=1}^{n}mke_{k-1,m}=mx+m\sum\limits_{k=1}^{n}e_{k-1,m}=n+x+ne_{n,m},\ed
which in view of Eq. \eqref{e1} gives Eq. \eqref{e2}.
Equation \eqref{e4} follows from Eqs. \eqref{e3} and \eqref{ek} as,
\bd
\begin{array}{rcl}
m^{k}\int\limits_0^{1/m} y^k(1-y)^{-2}\ud y&=&m^{k}\int\limits_0^{1/m} \left({m\over m-1}-{1\over1-y}\right)\ud y^k\\
&=&{m\over m-1}-km^{k}\int\limits_0^{1/m}{y^{k-1}\ud y\over1-y}.
\end{array}
\ed

{\sc Proof of Eqs. \eqref{cf} and \eqref{cf2}.} Let us write $o_m$ instead of $O\left((m-1)^2\ln {m\over m-1}\right)$ as $m\downarrow1$.
Observe that,  
\bd
a_n+e_{n,m}-\ln {m\over m-1}=\sum\limits_{i=1}^{n}\left({1\over i}-{m^{n-i}\over i}\right)+(m^n-1)\ln {m\over m-1}.
\ed
Thus as $m\downarrow1$,
\bd
\begin{array}{rcl}
{a_n+e_{n,m}-\ln {m\over m-1}\over m-1}&=&\left(\sum\limits_{i=0}^{n-1}m^{i}\right)\ln {m\over m-1}-\sum\limits_{i=1}^{n}{\sum\limits_{j=0}^{n-i-1}m^{j}\over i}\\
&=&n\ln {m\over m-1}+\ln {m\over m-1}\left(\sum\limits_{i=1}^{n-1}i\right)(m-1)\\
&&-\sum\limits_{i=1}^{n}\left({n-i\over i}+{(m-1)\sum\limits_{j=1}^{n-i-1}j\over i}\right)+o_m\\
&=&n\left(1-a_n+\ln {m\over m-1}\right)\\
&&+{n\choose2}\left(\ln {m\over m-1}-a_n+{3\over 2}\right)(m-1)+o_m,
\end{array}
\ed
since,
\bd
\begin{array}{rcl}
\sum\limits_{i=1}^{n}{\sum\limits_{j=1}^{n-i-1}j\over i}&=&\sum\limits_{i=1}^{n-2}{(n-i-1)(n-i)\over 2i}\\
&=&{n\choose2}a_{n-2}-{(2n-1)(n-2)\over 2}+{(n-1)(n-2)\over 4}={n\choose2}\left(a_{n}-{3\over 2}\right).
\end{array}
\ed
It follows,
\bd
\begin{array}{rcl}
1+e_{n,m}&=&\left(1-a_n+\ln {m\over m-1}\right)(1+n(m-1))+o_m,\\
n(1+e_{n,m})&=&\left(a_n+e_{n,m}-\ln {m\over m-1}\right)\left({1\over m-1}+n\right)\\
&&-{n\choose2}\left(\ln {m\over m-1}-a_n+{3\over 2}\right)(m-1)+o_m,
\end{array}
\ed
and 
\bd
\begin{array}{cl}
&{2(1+e_{n,m})\over (n-1)(m-1)}-{2m\left(a_n+e_{n,m}-\ln {m\over m-1}\right)\over n(n-1)(m-1)^2}\\
=&2\left(1-a_n+\ln {m\over m-1}\right)-\ln {m\over m-1}+a_n-{3\over 2}+{o_m\over m-1}\\
=&\ln {m\over m-1}-a_n+{1\over 2}+{o_m\over m-1}.
\end{array}
\ed
Combining these results we find that Eq. \eqref{cf1} indeed implies Eq. \eqref{cf2}:
\bd
\begin{array}{rcl}
\rho_n&=&2-{1\over e_{n,m}}\left({2(1+e_{n,m})\over (n-1)(m-1)}-{2m\left(a_n+e_{n,m}-\ln {m\over m-1}\right)\over n(n-1)(m-1)^2}\right)\\
&=&2-{\ln {m\over m-1}-a_n+{1\over 2}+{o_m\over m-1}\over\ln {m\over m-1}-a_n+{o_m\over m-1}}\\
&=&1-{1/2\over\ln {m\over m-1}-a_n+o(1)}.
\end{array}
\ed

Equation \eqref{cf} is derived from Eq. \eqref{sf} in a similar way.\\

{\sc Proof of Eqs. \eqref{cfa} and \eqref{cfa1}.} Equations \eqref{cfa} and \eqref{cfa1} are easily obtained from Eqs.\eqref{sf} and \eqref{cf1} using the following integral approximation for the function in 
Eq. \eqref{enl},
\bd e_{n,m}=\sum\limits_{i=1}^{\infty}{1\over (n+i)m^{i}}\to \int\limits_0^\infty {e^{-x}\ud x\over\alpha+x},\ n\to\infty,\ n(m-1)\to\alpha\ed
for a given positive $\alpha$. The last convergence follows from a Riemann sum representation,
\bd \sum\limits_{i=1}^{\infty}{1\over (n+i)m^{i}}=\sum\limits_{i=1}^{\infty}\delta f\left(i\delta\right),\ed
where $\delta=m-1$ and 
\bd f(x)={m^{-x/(m-1)}\over n(m-1)+x}\to {e^{-x}\over\alpha+x}.\ed
We can recognize that $e_{n,m}$ converges to a transformation of the exponential integral namely,
\bd e_{n,m} \to e^{-\alpha} \int\limits_{\alpha}^{\infty} {e^{-x}\over x} \ud x,\ n\to\infty,\ n(m-1)\to\alpha.\ed
The previously presented formulae for $e_{n,m}$ are not suitable for numerically calculating its value
but Graham Jones pointed out in personal correspondence that by a change of variables
\be
e_{n,m} = \int\limits_{\ln (m-1)}^{\ln (m)} (m-e^{x}) \ud x
\ee
which is well suited for computation. Alternatively, as again pointed out by Graham Jones,
in Eq. \eqref{enl} one can directly bound the tail (sum of terms from some $K_{0}$) of the infinite series by 
$m^{1-K_{0}}/((K_{0}+n)(m-1))$.

\bibliography{SagitovBartoszek}
\bibliographystyle{plainnat}

\end{document}